\documentclass{article}
\usepackage{spconf,amsmath,graphicx}
\usepackage{algorithm,algorithmic}
\pdfoutput=1 

\def\vx{\mathbf{x}}

\title{Language-agnostic Multilingual Modeling}
\name{Arindrima Datta, Bhuvana Ramabhadran, Jesse Emond, Anjuli Kannan, Brian Roark}
\address{\{arindrimadatta, bhuv, emond, anjuli, roark\}@google.com\\
Google Inc,\\
NYC, USA}
\begin{document}
\ninept
\maketitle
\begin{abstract}
Multilingual Automated Speech Recognition (ASR) systems allow for the joint training of data-rich and data-scarce languages in a single model. This enables data and parameter sharing across languages, which is especially beneficial for the data-scarce languages. However, most state-of-the-art multilingual models require the encoding of language information and therefore are not as flexible or scalable when expanding to newer languages. Language-independent multilingual models help to address this issue, and are also better suited for multicultural societies where several languages are frequently used together (but often rendered with different writing systems).
In this paper, we propose a new approach to building a language-agnostic multilingual ASR system which transforms all languages to one writing system through a many-to-one transliteration transducer. Thus, similar sounding acoustics are mapped to a single, canonical target sequence of graphemes, effectively separating the modeling and rendering problems. We show with four Indic languages, namely, Hindi, Bengali, Tamil and Kannada, that the language-agnostic multilingual model achieves up to 10\% relative reduction in Word Error Rate (WER) over a language-dependent multilingual model.

\end{abstract}

\section{Introduction}
\label{sec:intro}

Multilingual automated speech recognition (ASR) models have been studied extensively, spanning both hybrid Hidden Markov Models/Neural Networks ~\cite{fugen2003efficient, thomas2012icassp, tuske2013icassp, sercu2017icassp, cui2017asru} and more recently, End-to-End (E2E) models~\cite{watanabe2017asru, karafiat2018, rosenberg2017end}. All these models have one principle in common: they share data and parameters from all the languages they are trained on, allowing for robustness and better generalization. In doing so, a multilingual ASR system that is trained as a single model for all languages can benefit the data-scarce/low-resource languages by transferring the \emph{shared knowledge} to them. 

Prior work in training multilingual representations~\cite{ma12002multilingual, cutler2014language} and end-to-end models~\cite{watanabe2017language, kannan2019large} have demonstrated that the best performing models require conditioning on language information. This information can be used to track language switches within an utterance~\cite{seki2018end,waters2019language}, adjust language sampling ratios, or add additional parameters based on the data distribution~\cite{kannan2019large}. However, dependency on language information limits the ability of a multilingual model to be extended to newer languages.
For Indic languages, there are additional challenges from \emph{code-switching} in conversation: there is a considerable amount of variability in the usage of a second language (typically English) alongside native languages such as Tamil, Bengali or Hindi. This makes it challenging to model the context under which code switching occurs, and the language to which a spoken word should be assigned. The problem is compounded by inconsistent transcriptions and text normalization~\cite{vu2012first, emond2018transliteration, srivastava2018homophone}.

The Indic languages considered in this study overlap in acoustic and lexical content, due to either language family relations or the geographic and cultural proximity of the native speakers. However, their writing systems occupy different unicode blocks. This causes inconsistency in transcriptions: a common word, wordpiece or phoneme can be realized with multiple variants in the native-language writing systems, leading to increased confusions and inefficiency in data sharing when training a multilingual model. We propose a training strategy that maps all languages to one writing system through a many-to-one transliteration transducer. We show with Indic languages, that such a multilingual, end-to-end ASR system can outperform a multilingual ASR system conditioned on the language, particularly for the data-scarce languages. 

Balancing skewed distributions of data across languages, arising from variability in the number of native speakers of a language plays a key role in the performance of multilingual models. To address this, approaches that represent the world's languages as points in a language space~\cite{ragni2015language}, to enable efficient bootstrapping, sharing of parameters (particularly for Indic languages) and adaptation of speech recognition systems to any language, have been proposed. Techniques originally proposed for adaptation of models to speakers, domains and languages have also been successfully extended to address the difficulties of data imbalance in multilingual models \cite{kannan2019large,Swietojanski2014slt, tong2017multilingual, karafiat2014but}. 

In this work, we propose a strategy to balance the data effectively, making it suitable for language-agnostic multilingual training. We present a single, multilingual ASR model that is language-independent, yet achieves the same state-of-the-art performance as a model conditioned on the language information. It also addresses the challenges of code-switching, making it readily useful for real-world scenarios. To the best of our knowledge, this is the first work that demonstrates state-of-the-art performance with a multilingual ASR system that uses a simple, data normalization scheme to eliminate the need for language-dependence. 

The rest of the paper is organized as follows. We present our proposed method in Section \ref{sec:data_prep}, followed by a description of our model architecture in Section \ref{sec:methods}, and results in Section \ref{sec:results}. A detailed analysis of our results compared to schemes that use language-conditioning is presented in Section~\ref{ssec:la_vs_ld} and key insights are highlighted in Section~\ref{sec:conclusions}.

\section{Multilingual Data Processing: Proposed method }
\label{sec:data_prep}

\subsection{Script normalization using transliteration}
\label{ssec:transliteration}
Transliteration is a sequence-to-sequence mapping problem that aims to convert text from one writing system  to another. Since transliterating texts from Indic languages' native scripts to the target script of Latin has been effective in \cite{emond2018transliteration, hellsten2017transliterated}, we chose Latin as the common script to normalize all the training data.

Following \cite{emond2018transliteration, hellsten2017transliterated}, we make use of pair language models for transliteration. Also known as joint multi-gram models, these were first proposed in \cite{bisani2008joint} for grapheme-to-phoneme conversion.  These are n-gram models over ``pair'' symbols consisting of an input unicode codepoint paired with an output unicode codepoint, e.g., x:Y.  As with grapheme-to-phoneme conversion, given an input lexicon consisting of native script words and Latin script realizations of those words (known as romanizations), expectation maximization is used to derive pairwise alignments between symbols in both the native and Latin scripts.\footnote{Training lexicons for all languages investigated were annotated by native speakers. See \cite{emond2018transliteration} for further details on how they were collected.}  These symbol-aligned sequences are then used to train an n-gram model, in our case a 6-gram model.  An n-gram model over pair symbol sequences is a joint model over the input/output sequences, which can thus be used to transliterate in either direction.  The n-gram models are straightforwardly represented as weighted finite state transducers (WFSTs), either as an acceptor over pair sequences, or as a transducer, after separating the pair symbols into input and output symbols in the transducer.  Inverting the WFST swaps the input and output labels, so that the same joint model can be used to map from the Latin script to the specific native script or the other way around, providing the means for transliterating in either direction.  Further details on such methods can be found in \cite{emond2018transliteration}.



\subsection{Effect of script normalization}
\label{ssec:transliterationeffect}


To understand the extent to which these languages overlap, we studied phoneme-grapheme maps and words shared across these languages.  
Both these factors would impact the performance of a multilingual model by allowing for the grapheme from the wrong language to be output when the model is not conditioned on language information. For example, the graphemes
\includegraphics[trim=20 120 20 120, height=\fontcharht\font`\B,clip,]{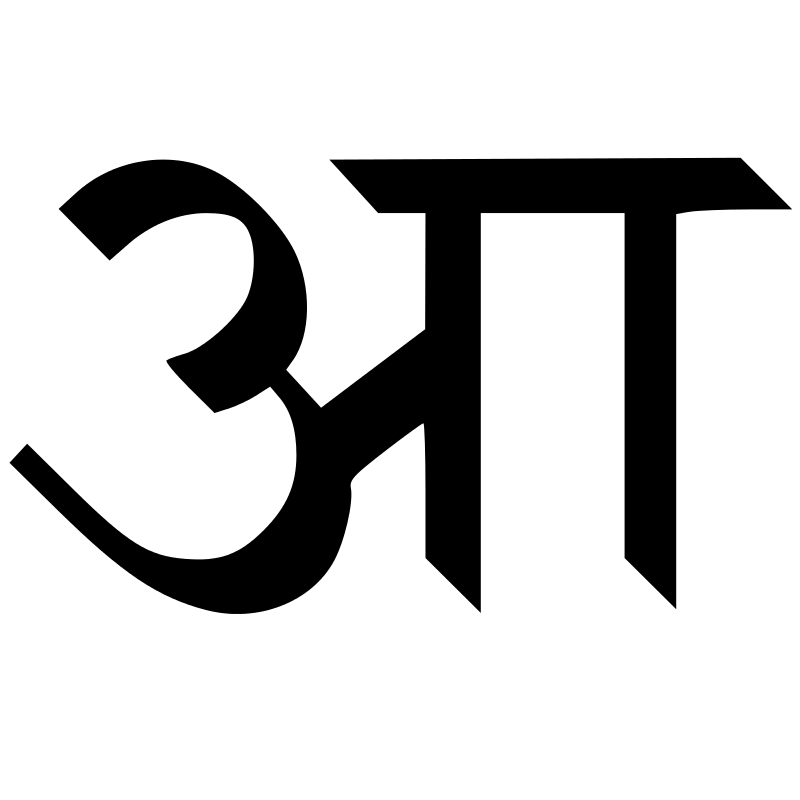},
\includegraphics[trim=20 120 20 140, height=\fontcharht\font`\B,clip]{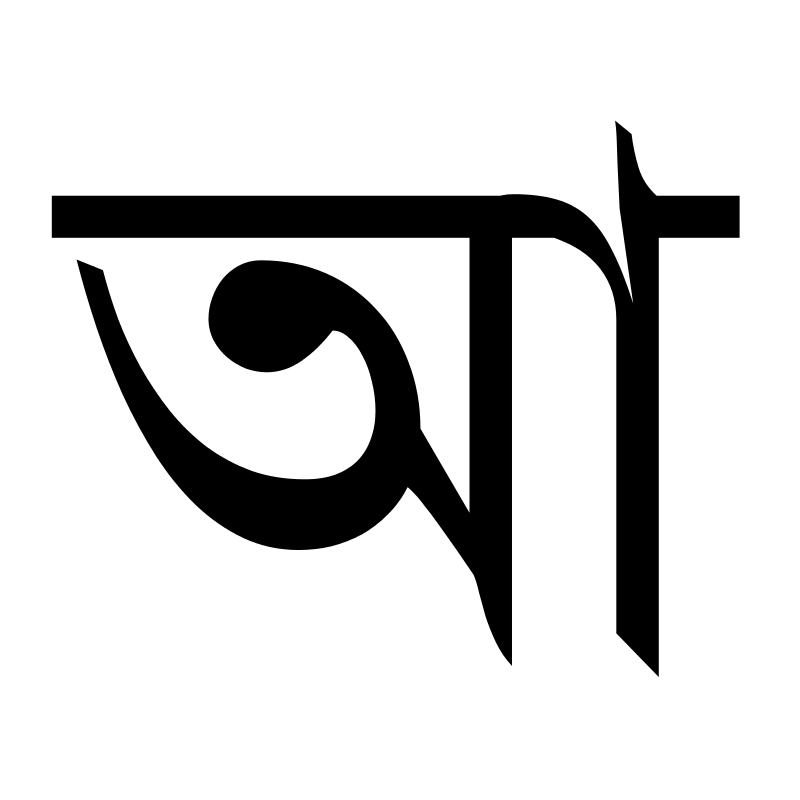},
\includegraphics[trim=20 0 20 0, height=\fontcharht\font`\B,clip]{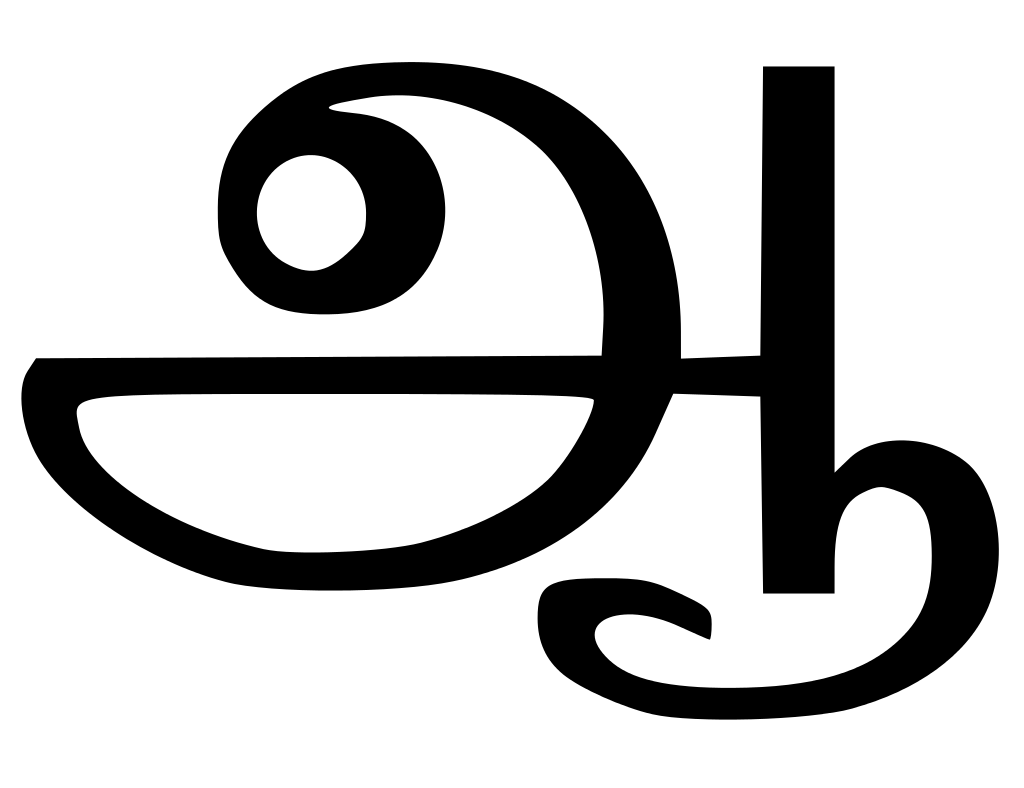},
\includegraphics[trim=0 10 0 10, height=\fontcharht\font`\B,clip]{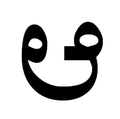}
in Hindi, Bengali, Tamil and Kannada map to the same phonetic sound, \emph{aa}. This implies that the multilingual model could hypothesize all four graphemes with different likelihoods if it were not conditioned on any language information. If all these graphemes were mapped to one canonical writing system, Latin, it would allow the model to share parameters for \emph{aa}. 
Table~\ref{fig:discovery_orig} illustrates the various possible romanizations of overlapping words in the training data, making the task of script normalization difficult.

\subsection{Proposed Transliteration Transducer}
\label{ssec:translit_fixes}

As stated in Section \ref{ssec:transliteration}, the input to training the pair language model transducers is a lexicon consisting of native script words and possible Latin script romanizations.  Importantly, there is no standard orthography in the Latin script in these languages, so that words can be spelled in a variety of ways.  Table \ref{fig:discovery_orig} shows native script spellings of the English word ``discovery'' in each of the four languages, along with attested romanizations of that word in the training data.  In all four languages, the actual spelling of the word in English is attested, however the annotators in each language may vary in the number and kind of romanizations they suggest, which may be driven by many factors, including differences in pronunciation or simply individual variation.


The variability of the training data in these languages can lead to an inconsistency in the mapping of shared vocabulary to a similar romanization. When a multilingual ASR model is trained with transliterated data, spelling inconsistency across languages creates a source of confusion and reduces the intended sharing of knowledge across the languages. Ideally, we would like English words, for example, to be mapped onto the same Latin script transcription of the word regardless of the native script in which they are originally written.
To mitigate such data inconsistencies, we propose two approaches for pre-processing the data for transliteration WFST training.

\begin{table}
\centering
\includegraphics[width=0.7\linewidth]{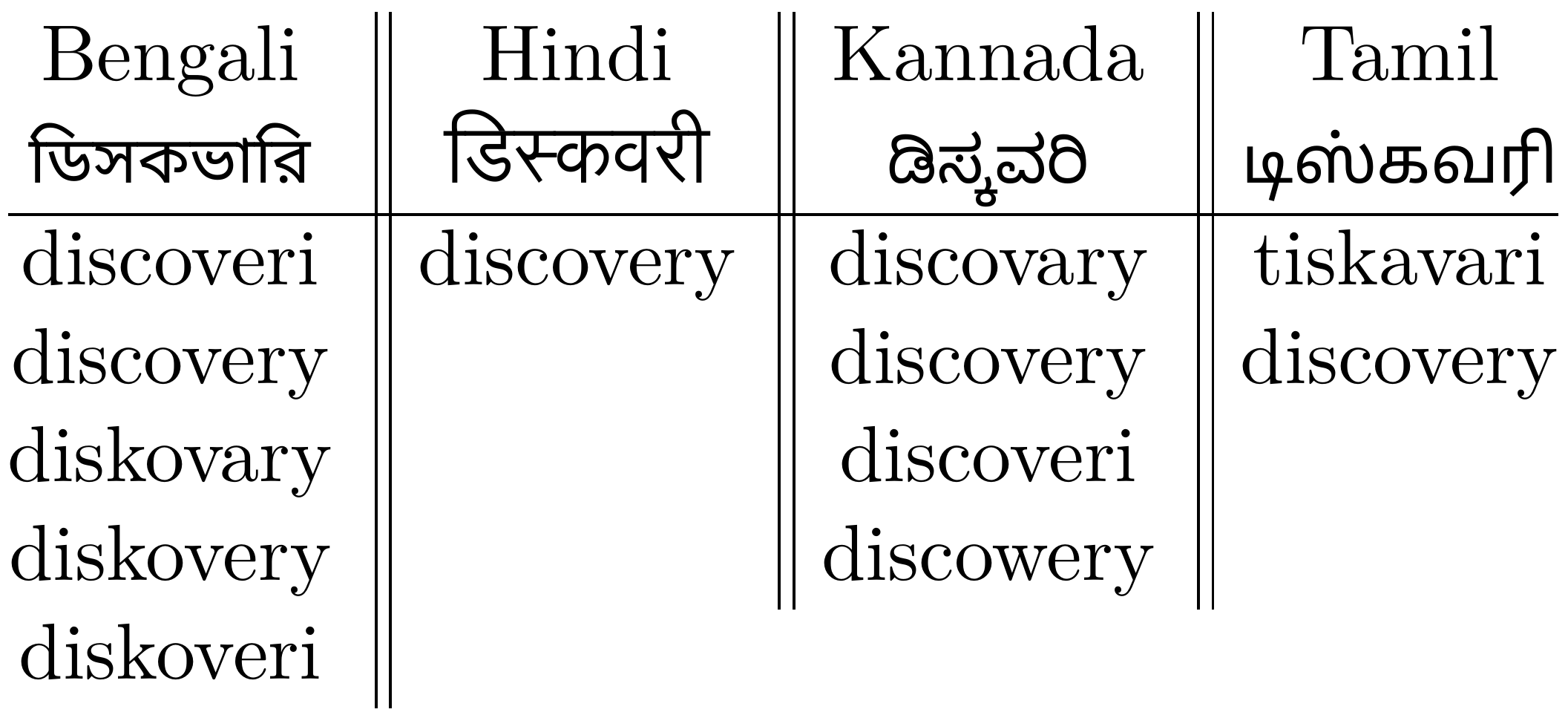}\vspace*{-0.05in}
\caption{\small Attested romanizations of the English word ``discovery'' in each of the four languages, illustrating the variation that can occur in transliteration training data.}\vspace*{-0.05in}
\label{fig:discovery_orig}
\end{table}

\subsubsection{Agreement-based (AB) Data Pre-processing}
\label{sssec:agreement_based}

In this method, we pre-process \emph{only the transliteration pairs which have at least one common transliterated form in all the four languages}. Native words that do not share a transliterated form are left unprocessed. As an example, the transliterated forms of the native words in Table~\ref{fig:discovery_orig} would only consist of {\bf discovery} after pre-processing.

\begin{algorithm}
\caption{Agreement-based pre-processing}
\begin{algorithmic}[AB]
  \scriptsize
  \STATE \textbf{HiWords}: Mapping from native Hindi words to Latin transliterated forms;
  \STATE \textbf{BnWords}: Mapping from native Bengali words to Latin transliterated forms;
  \STATE \textbf{TaWords}: Mapping from native Tamil words to Latin transliterated forms;
  \STATE \textbf{KnWords}: Mapping from native Kannada words to Latin transliterated forms;
  \STATE
  \STATE $common\_latin\gets$ Latin(HiWords) $\cap$ Latin(BnWords) $\cap$ Latin(TaWords) 
  \STATE $\cap$ Latin(KnWords)
  \STATE
  \FORALL{mapping in \{HiWords, BnWords, TaWords, KnWords\}}
    \FORALL{native\_word in Native(mapping)}
      \STATE $agreed\_latin\gets$ mapping[native\_word] $\cap$ common\_latin
      \IF{$agreed\_latin\neq\emptyset$}
        \STATE mapping[native\_word] $\gets agreed\_latin$
      \ENDIF
    \ENDFOR
  \ENDFOR
\end{algorithmic}
\end{algorithm}

\subsubsection{Frequency-based (FB) Data Pre-processing}
\label{sssec:frequency_based}

In addition to the native-transliterated word pairs, the training data also contained the frequency of occurrences of all transliterated forms for a native word.
Utilizing the frequencies in this pre-processing method we transformed \emph{ALL} transliteration pairs for each language. This approach is also based on our empirical observation that the most frequent transliterated forms were \emph{usually} the best, for instance, the commonly used spellings of proper nouns, or the dictionary spellings of the English words. Thus, for a native word, we retain only the transliterated forms that meet a frequency threshold (in this case, the average transliteration frequency per native word) and discard the rest.

\begin{algorithm}
\caption{Frequency-based pre-processing}
\begin{algorithmic}[FB]
  \scriptsize
   \STATE \textbf{Mappings}: For each language, mapping from native words to transliterated forms
   \FORALL{mapping in Mappings}
        \FORALL{native\_word in Native(mapping)}
            \STATE $translits\gets$ mapping[word]
            \STATE $avg\_freq\gets \frac{1}{| translits |} \sum_{t \in tranFreq}(t)$
            \STATE mapping[native\_word] $\gets \{t| t\in translits, Freq(t) \geq avg,freq\}$
        \ENDFOR
    \ENDFOR
\end{algorithmic}
\end{algorithm}

\subsection{Data Balancing}
\label{ssec:data_balancing}
Due to the varying distribution of language speakers and maturity of speech products, our training data is steeply skewed across languages. In our setup, the highest-resource language (Hindi) has two orders of magnitude more data than the lowest-resource language, Kannada. Any neural network trained on such imbalanced data tends to be more influenced by the over-represented languages in the training set~\cite{kannan2019large}. The effect is even more pronounced for end-to-end model architectures that neither have language information encoded with the acoustic features, nor incorporate language models.
Thus, in a language-agnostic multilingual system, where the model is not aided with a language identifier, it is even more critical to address data imbalance. 

We balanced the training set across languages by augmenting the original data with diverse noise styles~\cite{Chanwoo17}. The exact amount of data augmentation needed for each language is determined  empirically, by oberserving that the single-language recognizer for the lowest-resource language, Kannada degrades in performance when trained on more than 75 noisy copies of the original data. The remaining languages were augmented with the needed number of noise styles to result in equal amounts of data for each of the four languages used in the multilingual models.



\section{Multilingual Model}
\label{sec:methods}

A low-latency, end-to-end, Recurrent Neural Network Transducers (RNN-T) architecture, originally proposed in~\cite{graves} is used for all the models in this paper.
The RNN-T models used here are similar to the ones used in~\cite{kannan2019large, he2019streaming} and are suitable for interactive applications that require streaming ASR. To summarize, the architecture consists of the following: an {\it encoder} of stacked LSTM layers that transforms a  sequence
    of $d$-dimensional feature vectors $\mathbf{x} =
    (\mathbf{x}_1, \mathbf{x}_2, \cdots, \mathbf{x}_T)$, where $\mathbf{x}_t \in \mathbf{R}^d$, at each time step to a higher-order feature
    representation, denoted by
    ${\mathbf{h}_1^{\text{enc}}}, \cdots, {\mathbf{h}_T^{\text{enc}}}$; 
an LSTM-based {\it decoder} that processes the
    sequence of non-blank, hypothesized graphemes, $y_0, \ldots, y_{u_{i-1}}$ into a representation
    ${\mathbf{h}_{u_{i}}^{\text{dec}}}$; 
and a {\it joint network} that combines these to predict a distribution over the next output grapheme,
    $P(y_{i} | \vx_1, \cdots, \vx_{t_i}, y_0, \ldots, y_{u_{i-1}})$. 
The input features are 80-dimensional log-mel features, computed over a 25ms window and shifted every 10ms, which are further stacked with 7 frames to the left and down-sampled to 30ms frame rate. The RNN-T model comprises of eight 2,048-dimensional LSTM layers in the encoder, and two 2,048-dimensional LSTM layers in the prediction network, each of which is followed by a 640-dimensional projection layer. The joint network is composed of 640 hidden units. The language-dependent (LD) multilingual system has 454 unified-grapheme targets while the language-agnostic (LA) system has 44 Latin grapheme targets (all languages are transliterated to the latin writing system). Each of the single language models was trained with a combined set of graphemes covering both, Latin and native writing systems.

All models were trained in Lingvo \cite{shen2019lingvo} on $4 \times 4$ Tensor Processing Units \cite{tpu} slices with a batch size of
4,096. The metric we report is \textit{transliteration-optimized WER} \cite{emond2018slt}. Following \cite{he2019streaming}, we also explored the use of a time-reduction layer with reduction factor N=2 after the second encoder layer of the RNN-T model in order to speed-up training and inference.


\section{Experiments, Results, And Analysis}
\label{sec:results}

\subsection{Data}
\label{ssec:data}
The training and test data for the four Indic languages consist of anonymized, human-transcribed utterances representative of Google's voice search traffic described in~\cite{kannan2019large, datta_kanna_2019}. Of the four languages, Hindi, Bengali, Tamil and Kannada, Hindi has the most amount of training data and Kannada (further reduced by a factor of six from what is described in~\cite{kannan2019large}) has nearly two orders of magnitude less training data than Hindi, with Tamil and Bengali falling in between. The training data is augmented with additional copies from noisy utterances created by corrupting the utterances using a room simulator and noise styles~\cite{Chanwoo17} .

\subsection{Results}
\label{ssec:wer}
In this Section, we describe the results from transilteration transducers trained with the methods proposed in Section~\ref{ssec:translit_fixes}. We present two sets of results. Unless otherwise noted, all Word Error Rates (WERs) are transliteration-optimized WERs introduced in~\cite{emond2018transliteration}. First, we validate our approach by evaluating these transducers 
for transliterated scoring~\cite{emond2018transliteration} only. Table~\ref{table:translit_fixes_sl} presents these results with the single language models. The ASR baseline for all languages, presented as S0 in Table~\ref{table:translit_fixes_sl} uses the transliteration transducer described in~\cite{emond2018transliteration}. Both proposed methods, {\it Agreement Based} and {\it Frequency Based} data pre-processing provide WER reductions over the baseline ASR model, (rows S1 and S2 in Table~\ref{table:translit_fixes_sl}).
The improvements from the two data pre-processing methods are similar 
except for Hindi. The training data for Hindi contains several valid romanizations occurring with similar frequencies, and {\it Frequency Based} pre-processing eliminated some of these. We hypothesize that this resulted in transliteration errors when rendering ASR hypotheses in the native writing system. The frequency distributions of romanizations for the other languages are more skewed with many transliterated forms occurring with low frequencies (Figure~\ref{fig:discovery_orig}). Therefore, these other languages benefit from this average frequency based pruning of rare and potentially erroneous romanizations. On the other hand, {\it Agreement Based} data pre-processing selects transliteration variants that are common across languages, and affects only words with at least one such common transliteration from all the languages; typically proper nouns or borrowed English words. Therefore, this results in consistent transilteration and rendering in the native writing systems.

Next, having established the validity of the proposed method, we study the impact on multilingual models both in transliteration of training data and in transliterated scoring. A similar trend in WER reductions is seen in Table~\ref{table:translit_fixes_ml}. Here, we observe that for Hindi, the performance gap between the {\it Agreement Based} and {\it Frequency Based} data pre-processing methods is further reduced. This can be attributed to the fact that an increased number of valid romanizations with higher frequency of occurrence can now be recovered from the other languages.

\begin{table}[!t]
  \centering
  \begin{tabular}{|p{0.8cm}|p{2.3cm}||p{0.5cm}|p{0.5cm}|p{0.5cm}|p{0.5cm}||p{0.5cm}|} \hline
  Exp  & Model               & Hi   & Bn   & Ta   & Kn   & Avg \\ \hline
    S0 & Baseline           & 19.0 & 20.4 & 28.8 & 50.2 & 29.6   \\
    S1 & S0 + AB            & 18.9 & 19.6 & 26.7 & 49.4 & 28.6   \\
    S2 & S0 + FB            & 19.3 & 19.6 & 26.9 & 49.4 & 28.8   \\ \hline
  \end{tabular}
  \caption{Effect of the proposed Agreement-based (AB) and Frequency-based (FB) data processing approaches on Word Error Rate of single-language models.}
  \label{table:translit_fixes_sl}
\end{table}

\begin{table}[!t]
  \centering
  \begin{tabular}{|p{0.8cm}|p{2.3cm}||p{0.5cm}|p{0.5cm}|p{0.5cm}|p{0.5cm}||p{0.5cm}|} \hline
  Exp  & Model               & Hi   & Bn   & Ta   & Kn   & Avg \\ \hline
    M0 & Baseline  & 22.8 & 24.2 & 30.5 & 36.7 & 28.6   \\
    M1 & M0 + AB  & 22.9 & 22.4 & 26.8 & 32.6  & 26.2   \\
    M2 & M0 + FB  & 22.9 & 22.2 & 27.6 & 33.1 & 26.4   \\ \hline
  \end{tabular}
  \caption{Effect of the proposed Agreement-based (AB) and Frequency-based (FB) data processing approaches on Word Error Rate of multilingual models.}
  \label{table:translit_fixes_ml}
\end{table}

\subsection{Language-agnostic vs Language-dependent models}
\label{ssec:la_vs_ld}

In this Section, we present the results from the best performing language-dependent and language-agnostic models. Our ablation studies presented in the earlier sections were conducted with the model architecture described in Section~\ref{sec:methods}. We found that the RNN-T architecture could be optimized further for both single-language and multilingual models by removing the time-reduction layer (See Table~\ref{table:stacking_layer}) which was introduced in~\cite{he2019streaming} to reduce training time. 
\begin{table}[!t]
  \centering
  \begin{tabular}{|p{0.8cm}|p{2.3cm}||p{0.5cm}|p{0.5cm}|p{0.5cm}|p{0.5cm}||p{0.5cm}|} \hline
  Exp  & Model               & Hi   & Bn   & Ta   & Kn   & Avg \\ \hline
    S1   & S0 + AB & 18.9 & 19.6 & 26.7 & 49.4 & 28.6   \\
    S3   & S1 -- TR         & 18.5 & 17.4 & 25.0 & 44.9 & 26.4   \\ \hline
    M1   & M0 + AB  & 22.9 & 22.4 & 26.8 & 32.6 & 26.2   \\
    M3   & M1 -- TR         & 21.6 & 20.8 & 25.6 & 30.5 & 24.6   \\ \hline
  \end{tabular}
\caption{Performance (WER) of Single-language and Multlingual models with and without time-reduction (TR) layers.}
\label{table:stacking_layer}
\end{table}

\begin{table}[!t]
  \centering
  \begin{tabular}{|p{0.3cm}|p{3.2cm}||p{0.45cm}|p{0.45cm}|p{0.45cm}|p{0.45cm}||p{0.45cm}|} \hline
    Exp  & Model                & Hi   & Bn   & Ta   & Kn   & Avg \\ \hline
    LD   & Language-dependent & 21.7 & 23.1 & 27.1 & 33.1 & 26.2   \\ \hline
    LA   & Language-agnostic (M3) & 21.6 & 20.8 & 25.6 & 30.5 & 24.6   \\ \hline
  \end{tabular}
\caption{Performance (WER) of language-agnostic (LA) and language-dependent (LD) multilingual models in comparison to the single language models}
\label{table:multilang_comparison}
\end{table}

Using the best data processing strategy from Section \ref{ssec:translit_fixes}, and the best performing RNN-T architecture (Table~\ref{table:stacking_layer}), we trained a multilingual model for the four languages, and compared its performance with a language-dependent (LD) model. Both models use the same architecture and training data, except that the language-dependent model hypothesizes from a unified set of graphemes across the four languages and uses language information in the first encoder layer \cite{grace, li}. In Table~\ref{table:multilang_comparison}, we see that the language-agnostic (LA) model is competitive in performance to the language-dependent (LD) model, beating the performance of the language-dependent model for 
all languages.

\begin{figure}
    \centering
  \includegraphics[height=3.9cm, trim=50 70 0 66,clip]{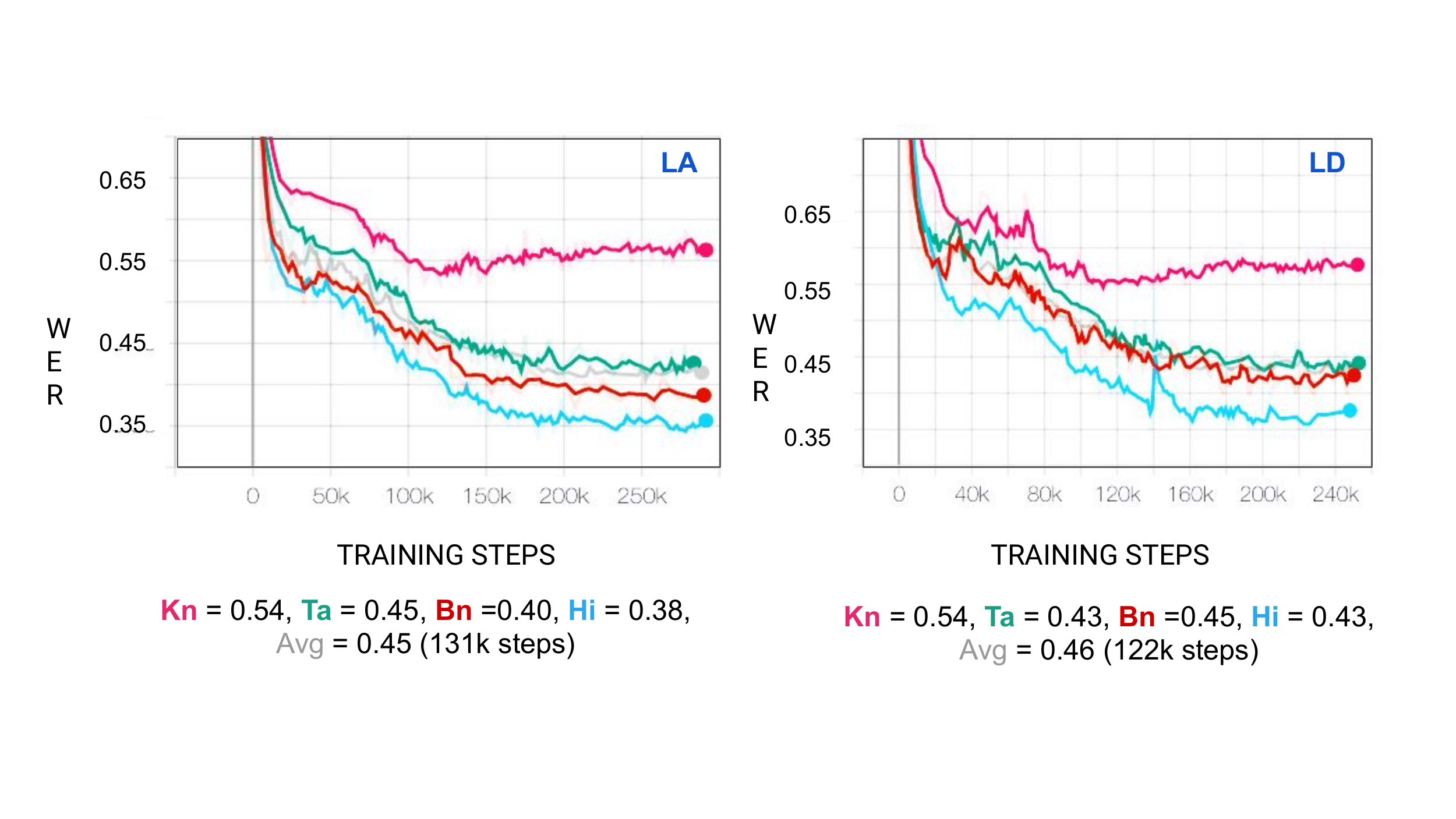}
\caption{Untransliterated Word Error Rate plots comparing language-agnostic [left] , and language-dependent [right] multilingual models on held out development test sets for all languages. WERs at convergence are noted below each plot for all languages.}
\label{fig:la_vs_ld}
\end{figure}

In multilingual models, languages compete for capacity given the
limited model size, training strategy and stopping criterion. Figure~\ref{fig:la_vs_ld} illustrates the convergence and performance properties of the language-dependent and language-agnostic multilingual models on a held-out development set. We used average WER computed on the development test set over all languages as the metric to decide the point of convergence. We observed that 
at approximately 130K steps, both the LD and LA models begin to overfit on the data-scarce language, Kannada (depicted by the magenta curve in the figure) while continuing to improve in performance over the other languages. This suggests that the models still had not converged for the data-rich languages, i.e, had not been trained 
on significant portions of the data-rich languages. However, the LA model yielded a much lower WER on the data-rich language, Hindi (depicted by the blue curve in the figure) without degrading the performance on the other languages, suggesting better data and parameter sharing.


\section{Conclusions}
\label{sec:conclusions}

We have proposed a new approach to building language-agnostic multilingual ASR systems for Indic languages by transforming data from multiple languages to one writing system through a many-to-one transliteration transducer. This approach maps similar sounding  acoustics to a single, canonical target sequence of graphemes, effectively separating the modeling and rendering problems. We show with four Indic languages, Hindi, Bengali, Tamil and Kannada, that the proposed language-agnostic multilingual model can provide an average reduction in WER of 6\% relative over the language-dependent multilingual model.
Relative to the LD model, the LA model also reduces WER on the language with the least amount of data, Kannada by 8\%, Bengali by 10\%, and Tamil by 6\% relative, while matching the performance on the most data-rich language, Hindi. The LA model offers other benefits such as the flexibility to scale to new languages, as well as its inherent ability to handle code-switching that is common among speakers of multilingual societies. Finally, the reduced number of modeling units resulting from the use of one canonical writing system (Latin) allows the LA model to be trained on other modeling units such as wordpieces that have been proven to provide performance wins over graphemes for most languages. 


\vspace{0.2cm}
\noindent
\textbf{Acknowledgements}: The authors would like to thank Parisa Haghani, Neeraj Gaur, Eugene Weinstein, Pedro Moreno and Meysam Bastani for helpful discussions and support.

\bibliographystyle{IEEEbib}
\bibliography{refs}

\end{document}